\documentstyle[epsfig,twoside,fleqn,espcrc2]{article}

\newcommand{\AmS}{{\protect\the\textfont2
  A\kern-.1667em\lower.5ex\hbox{M}\kern-.125emS}}

\title{Quantum fluctuations of the color flux tube}

\author{P. Provero\address{
Dipartimento di Fisica Teorica, Universit\`a di Torino,\\
Via P. Giuria 1, 10125 Torino, Italy\\
e--mail: provero@to.infn.it}\thanks{Work done in collaboration with
M. Caselle, R. Fiore, F. Gliozzi and M. Hasenbusch}}%

\begin{document}

\begin{abstract}
The quantum fluctuations of the flux tube joining two static sources in the
confining phase of a lattice gauge theory are described by an effective string
theory. The predictions of the latter for ratios of Wilson loops of equal
perimeter do not contain any free parameters, and can be computed exactly for
large Wilson loops. We compare these predictions with numerical results in 3D
$Z_2$ gauge theory, finding complete agreement.
\end{abstract}

\maketitle

\section{INTRODUCTION}

The flux--tube picture of confinement is now a 20 year old conjecture, 
which we can summarize in two statements:
\begin{itemize}
\item{Two static sources in the confined phase of a lattice gauge theory are
joined by a color flux tube which is responsible for confinement.}
\item{The flux tube can fluctuate around its equilibrium position; in the rough
phase, these fluctuations are massless \cite{luscher}.}
\end{itemize}
These two hypotheses not only provide an intuitive, physical picture of
confinement, but have precise quantitative consequences. In principle, these
predictions can be compared with the experimental data for the spectrum of bound
states of heavy quarks; in practice, precise tests are currently possible using
numerical simulations.\\
The aim of this work is to show that the precision now available in Monte Carlo
data for the simplest lattice gauge theory, namely the 3D $Z_2$ gauge model,
allows one to say a final word about the correctness of the fluctuating flux
tube picture.
\section{THE PREDICTIONS OF THE FLUCTUATING FLUX TUBE MODEL}
The natural quantity to be studied in this context is the Wilson loop $W(R,T)$.
Suppose two static sources are pulled apart at a distance $R$, kept apart for a
(euclidean) time $T$, then put back together. If we assume that the flux tube
keeps its equilibrium position, a straight line joining the two sources, the
expectation value of the Wilson loop will follow the area law:
\begin{equation}
\langle W(R,T)\rangle\propto e^{-\sigma RT}\label{area}
\end{equation}
where $\sigma$ is the string tension. For the interquark potential this
corresponds to a linear confining potential:
\begin{equation}
V(R)=\sigma R
\end{equation}
If, instead, we allow the flux tube to fluctuate around its equilibrium
position, we obtain corrections to the area law (\ref{area}) that can be
compared to the results of Monte Carlo simulations. 
Many different models
can be written down to describe the dynamics of the flux tube fluctuations; 
however,
a vast class of these reduce to an effective 
free string model in the infrared region, {\em i.e.} for
large Wilson loop. In this region the expectation value of
the Wilson loop is given by
\begin{eqnarray}
&&\langle W(R,T)\rangle\propto e^{-\sigma RT}\cdot\nonumber\\
&&\ \ \int [dX^i]\exp\left\{-\frac{\sigma}
{2}\int d^2\xi X^i(-\partial^2)X^i\right\}\label{fstr}
\end{eqnarray}
The fields $X^i,\ (i=2,\dots,d-1)$ are defined on the rectangle $(0,R)\times
(0,T)$ and satisfy Dirichlet boundary conditions. They describe the fluctuations
of the flux tube in the $d-2$ directions transverse to the Wilson loop.
\\
The functional integral is gaussian and can be performed exactly,
using for example $\zeta$-function regularization. The result is \cite{det}
\begin{equation}
\langle W(R,T)\rangle\propto e^{-\sigma R T}\left[\frac{\eta(\tau)}
{\sqrt{R}}\right]^{-\frac{d-2}{2}}\label{pred}
\end{equation}
where $\tau=i\frac{T}{R}$ and $\eta(\tau)$ is the Dedekind function, which can
be expressed as an infinite product
\begin{equation}
\eta(\tau)=q^{1/24}\prod_{n=1}^{\infty}\left(1-q^n\right)
\end{equation}
in terms of the variable $q=e^{2\pi i\tau}$. Eq. (\ref{pred}) is the prediction
of the effective string model for the expectation value of large Wilson
loops. It
should be noted that that Eq. (\ref{pred}) does not contain any new adjustable
parameters with respect to the simple area law (\ref{area}).
\section{RATIOS OF WILSON LOOPS}
Our goal is to compare Eq.(\ref{pred}) with the results of Monte Carlo
simulations of the 3D $Z_2$ gauge model. The choice of the model is dictated by
the need to have very high precision data to obtain an unambiguous result about
the validity of the string model. This test could be performed
by fitting a set of Monte Carlo data for the Wilson loop with Eq.(\ref{pred}).
Taking into account the presence of a perimeter term, this would be a fit with
three adjustable parameters.\\
However, to obtain a completely unambiguous test of our ansatz, we decided to
follow a different strategy, namely to define an observable quantity for which
the predictions of the string model do not contain {\em any} adjustable
parameters. We considered the following ratios of Wilson loops:
\begin{equation}
R(L,n)=\frac{\langle W(L+n,L-n)\rangle}{\langle W(L,L)\rangle}e^{-\sigma n^2}
\label{defrln}
\end{equation}
\par
Using Eq.(\ref{pred}) we obtain the prediction of the effective free string
model for the ratios $R(L,n)$, which turns out to depend only on the asymmetry
ratio $t=n/L$:
\begin{equation}
R(L,n)=F(t)=\left[\frac{\eta(i)
\sqrt{1-t}}{\eta\left(i\frac{1+t}{1-t}\right)}\right]^{1/2}\label{ft}
\end{equation}
\section{COMPARISON TO THE MONTE CARLO DATA}
\begin{table*}[hbt]
\setlength{\tabcolsep}{1.5pc}
\newlength{\digitwidth} \settowidth{\digitwidth}{\rm 0}
\catcode`?=\active \def?{\kern\digitwidth}
\caption{Monte Carlo results for the ratios $R(L,n)$}
\label{tab:compari}
\begin{tabular*}{\textwidth}{@{}l@{\extracolsep{\fill}}rrllr}
\hline
$n/L$&$L$&$\beta$&$\sigma$&$R(L,n)$&$F(n/L)$\\
\hline
$0.2$   &$15$&$0.75202$&$0.01023$& $1.0104(17)$&$1.00881$\\
$0.25$  &$20$&$0.75632$&$0.004779$&$1.0166(10)$&$1.01453$\\
$0.33$  &$12$&$0.74883$&$0.014728$&$1.02881(30)$&$1.02901$\\
$0.375$ &$ 8$&$0.75245$&$0.009418$&$1.0403(31)$&$1.03940$\\
$0.45$  &$20$&$0.75632$&$0.004779$&$1.0684(23)$&$1.06588$\\
$0.5$   &$20$&$0.75632$&$0.004779$&$1.0911(27)$&$1.09153$\\
$0.6$   &$25$&$0.75632$&$0.004779$&$1.165(11) $&$1.17667$\\
\hline
\end{tabular*}
\end{table*}
\begin{figure}
\begin{center}
\mbox{\epsfig{file=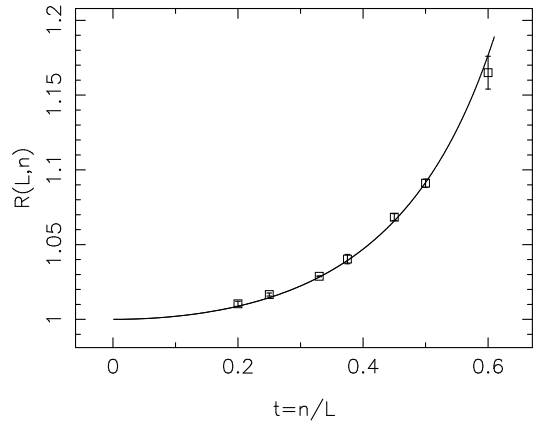}}
\vskip 2mm
\caption{Comparison of the predictions of the free string model with Monte Carlo
data for the ratios $R(L,n)$. The solid line is the prediction of the 
string model, Eq.(\ref{ft}).}
\label{fig:compa}
\end{center}
\end{figure}
We performed a Monte Carlo simulation of the 3D $Z_2$ gauge model at several
values of the coupling $\beta$, all located in the rough phase and close enough
to the deconfinement point to be well inside the scaling region.
From these simulations we extracted the expectation values of the Wilson loops.
To evaluate $R(L,n)$, one needs also to know the value of the string tension
$\sigma$. However, precise evaluations of the interface tension in the 3D {\em
spin} Ising model in the scaling region are available. Using
the duality between the $Z_2$ gauge model and the Ising spin model in 3D, we can
simply plug the Ising model values in the expression (\ref{defrln}). In this way
we use values of $\sigma$ that are independent from our Monte Carlo samples.
\par
The results of the comparison are shown in Fig.(\ref{fig:compa}) 
and Tab.(\ref{tab:compari}).
For each value of the
asymmetry ratio $t=n/L$ we have displayed the value of the ratio $R(L,n)$
with the largest available physical size $\sigma L^2$. The solid line is
the prediction, Eq. (\ref{pred}), of the effective free string model.
\begin{figure}
\begin{center}
\mbox{\epsfig{file=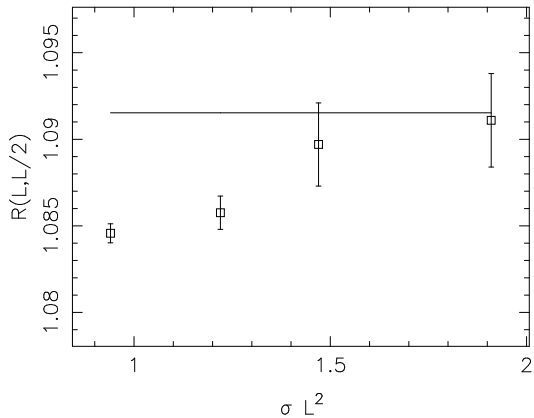}}
\vskip 2mm
\caption{Finite--size effects for small Wilson loop.
The prediction of the free string model is $R(L,L/2)=1.09153\dots$
(straight line).}
\label{fig:finite}
\end{center}
\end{figure}

Two conclusions can be drawn from these data:
\begin{itemize}
\item{The fluctuations of the flux tube are quantitatively relevant: it is easy
to see that neglecting them the ratios $R(L,n)$ would be predicted to be all
equal to one}
\item{The free string model, Eq.(\ref{fstr}), describes the fluctuations with
great accuracy for large Wilson loops}
\end{itemize}
\par
For Wilson loops smaller than a threshold size of order $\sigma L^2\sim 1$,
rather significant finite--size effects appear.
In Fig. (\ref{fig:finite})   we
display various data for ratios $R(L,L/2)$, {\em i.e.} with the same asymmetry
ratio $t=1/2$, and different sizes $\sigma L^2$.
\par
In conclusion, we have shown that, at large distances,
the physics of the flux tube fluctuations
is accurately described by a
free string model.
For shorter distances, finite--size effects are rather
significant. Precise data for large Wilson loops are needed to distinguish
between the free string model described here and other possible models.
For example, in Ref.\cite{ferm} a fermionic string model was shown to describe
the flux--tube fluctuations with an accuracy comparable to or
even greater than the free,
bosonic string model.
The high precision data now available for large Wilson
loops allow us to select the bosonic model as the most accurate.

\end{document}